\documentclass[10pt,twocolumn,showpacs,preprintnumbers,amsmath,amssymb,aps,prl,superscriptaddress,longbibliography]{revtex4-2}
\usepackage{appendix}
\usepackage[english]{babel}
\usepackage{bbm}
\usepackage{mathrsfs}
\usepackage{braket}
\usepackage{amsmath}
\usepackage{amsfonts}
\usepackage{CJKutf8}
\usepackage[dvipsnames]{xcolor}
\usepackage[colorlinks=true,linkcolor=Blue,urlcolor=BlueViolet,citecolor=BlueViolet]{hyperref}
\usepackage{natbib}
\usepackage{multirow}
\usepackage{graphicx}% Include figure files
\usepackage{dcolumn}% Align table columns on decimal point
\usepackage{bm}% bold math
\begin{document}

\title{\textbf{Machine learning reveals common features of unconventional superconductors with high transition temperatures} 
}% 

\author{Haosheng Xu}
 \thanks{These authors contribute equally to the work.}
 \affiliation{State Key Laboratory of Surface Physics and Department of Physics, Fudan University, Shanghai 200433, China}
\affiliation{Shanghai Research Center for Quantum Sciences, Shanghai 201315, China}

\author{Dongheng Qian}%
 \thanks{These authors contribute equally to the work.}
 \affiliation{State Key Laboratory of Surface Physics and Department of Physics, Fudan University, Shanghai 200433, China}
\affiliation{Shanghai Research Center for Quantum Sciences, Shanghai 201315, China}

\author{Yijun Yu}
  \affiliation{State Key Laboratory of Surface Physics and Department of Physics, Fudan University, Shanghai 200433, China}

\author{Jing Wang}
\thanks{wjingphys@fudan.edu.cn}
\affiliation{State Key Laboratory of Surface Physics and Department of Physics, Fudan University, Shanghai 200433, China}
\affiliation{Shanghai Research Center for Quantum Sciences, Shanghai 201315, China}
\affiliation{Institute for Nanoelectronic Devices and Quantum Computing, Fudan University, Shanghai 200433, China}
\affiliation{Hefei National Laboratory, Hefei 230088, China}

\date{\today}

\begin{abstract}
Superconductors with high critical temperatures that emerges beyond the phonon-mediated regime are usually considered unconventional in nature, yet unlike conventional superconductors, no broadly applicable predictive theory currently guides their discovery. Here, we use interpretable machine learning to uncover a common materials-space signature of high-$T_{\mathrm{c}}$ unconventional superconductors and develop a data-driven strategy for materials discovery. We construct a unified feature representation for each material by integrating compositional statistics, structural information, and latent representations from trained property-prediction models, followed by structure-aware filtering of an experimentally established superconducting dataset. Without using transition-temperature information, unsupervised analysis shows that cuprate and iron-based superconductors occupy a common region of materials space, characterized primarily by large electronegativity deviation and intermediate mean valence-electron number. A supervised $T_{\text{c}}$ model independently identifies the same descriptors as dominant features, providing complementary evidence for their relevance. Using this empirical materials-space prior together with the $T_{\text{c}}$ model, we prioritize candidate materials, recover recently discovered nickelate superconductors, and identify chemically distinct candidates for future investigation.
\end{abstract}

\maketitle

Superconductivity has long been a central topic in condensed-matter physics, combining fundamental interest in macroscopic quantum phenomena with broad technological applications. A central challenge is to understand why superconductivity emerges in chemically and structurally diverse materials and how microscopic characteristics determine their superconducting properties. Before a microscopic description of phonon-mediated superconductivity became available as a predictive framework for materials discovery, empirical rules had already played an important role in the search for superconductors. In the 1950s, Matthias and co-workers identified systematic correlations between superconductivity and materials characteristics, which were subsequently distilled into the ``Matthias' rules''~\cite{Matthias1963} and provided valuable guidance for the discovery of conventional superconductors with high $T_{\mathrm{c}}$. The microscopic description of phonon-mediated superconductivity was subsequently established within Bardeen--Cooper--Schrieffer (BCS) theory~\cite{Bardeen1957} and its Eliashberg extension~\cite{liashberg1960}, enabling quantitative understanding of conventional superconductivity, including the near-room-temperature hydrides under megabar pressures~\cite{Drozdov2015, Drozdov2019}. However, many superconducting families lie beyond this conventional framework. In particular, cuprates~\cite{Bednorz1986, Keimer2015} and iron-based compounds~\cite{Kamihara2008}, exhibit high transition temperatures at ambient pressure, yet a broadly applicable theory that describes their pairing mechanism and provides predictive guidance for their discovery remains unavailable. Consequently, the discovery of unconventional superconductors has relied largely on experimental exploration and empirical knowledge accumulated across individual material families. Therefore, establishing systematic criteria that can efficiently guide the search for new high-temperature superconductors remains a central open challenge.

Artificial intelligence (AI) and machine learning have recently emerged as powerful tools for materials discovery~\cite{Cheng2026}, with applications ranging from property prediction~\cite{Xie2018, Choudhary2021, xu2025} and crystal structure generation~\cite{Zeni2025, jiao2024space} to inverse design of materials with desirable properties~\cite{REN2022314, Park2026, xu2024, xu2026, Okabe2026}. For conventional superconductors, the availability of the electron--phonon framework provides a well-established theoretical foundation for quantitative prediction and has facilitated AI-driven discovery. Machine-learning models have been used to predict transition temperatures and learn physically meaningful quantities such as the Eliashberg spectral function $\alpha^2F(\omega)$~\cite{Choudhary_Kamal2022, Gibson2025, Gibson2026}, while comprehensive workflows combining generative models, active learning, and fine-tuning have also been developed~\cite{Wines_Daniel2023, XiaoQi2025, Prakash2026, MingzeLi2026}. Extending such strategies to unconventional superconductors is more challenging in the absence of a comparable predictive theory. Nevertheless, databases combining experimentally measured transition temperatures with crystal-structure information have recently become available for unconventional superconductors~\cite{SuperCon2022, Sommer2023, Konno2021, Pereti2023, Stanev2018, Hamidieh2018, Court2020}, providing an opportunity to identify common materials-space features directly from data, as what Matthias and his colleagues did in 1950s. Previous studies have proposed organizing principles for unconventional superconductivity, including the possible role of $\sigma$-bond metallization~\cite{An2001, Gao2015, Gao2021} and shared structural and electronic characteristics across cuprate and iron-based families~\cite{Hu2015, Hu2016, Le2018}. While these studies provide valuable physical insight, many such proposals are family-specific or rely strongly on human intuition. This raises a natural question: can data-driven analysis systematically identify simple, transferable materials-space characteristics shared by high-$T_{\mathrm{c}}$ unconventional superconductors, without assuming a particular microscopic pairing mechanism, and use them to prioritize materials for further investigation~\cite{Omri2026}?

\begin{figure*}[t]
\begin{center}
\includegraphics[width=5.5in, clip=true]{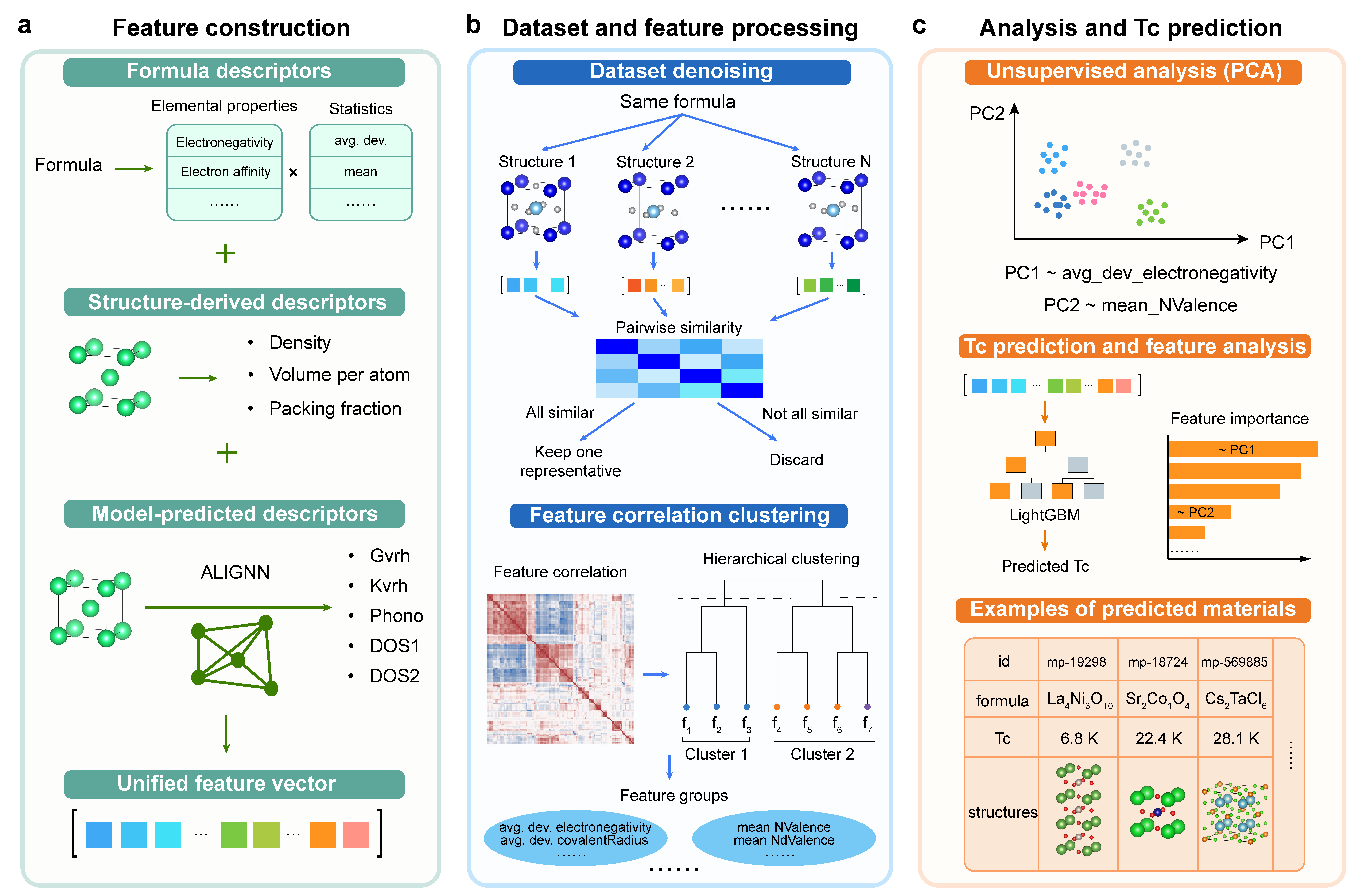}
\end{center}
\caption{\textbf{Overview of the machine-learning workflow.} \textbf{a}, Construction of a unified 147-dimensional feature vector from composition-derived descriptors, structure-derived descriptors, and latent representations extracted from pre-trained ALIGNN models. \textbf{b}, Dataset curation through structure-aware filtering, and organization of the descriptors into feature groups by hierarchical clustering. \textbf{c}, Unsupervised PCA analysis, supervised $T_{\mathrm{c}}$ prediction with feature importance analysis, and candidate screening.}
\label{fig1}
\end{figure*}

Here, we identify a common materials-space feature of high-$T_{\text{c}}$ unconventional superconductors using interpretable machine learning and demonstrate how this feature can be used to prioritize candidate materials. The overall workflow is illustrated in Fig.~\ref{fig1}. We construct an integrated representation of experimentally established superconductors by combining compositional statistics, structural information, and latent representations from a trained atomistic model. Unsupervised learning is then used to identify characteristic regions of materials space associated with unconventional superconductors, whereas supervised learning is used to model the dependence of $T_{\mathrm{c}}$ on the underlying descriptors. Our analysis reveals that cuprate, iron-based, and nickelate superconductors, despite their distinct crystal structures and chemical compositions, occupy an overlapping region of materials space. Remarkably, this region is primarily characterized by two simple and physically interpretable descriptors: a large average deviation of electronegativity and an intermediate mean number of valence electrons (Fig.~\ref{fig1}c). The correlation with superconducting temperature is systematic: the upper envelope of observed $T_{\text{c}}$ increases with average electronegativity deviation, a trend that persists within individual iron-based subfamilies, while high-$T_{\text{c}}$ materials consistently occur within an intermediate range of mean valence-electron number. A supervised model trained to predict $T_{\text{c}}$ independently identifies the same descriptors as dominant features, providing complementary evidence that these characteristics capture information relevant to high-$T_{\text{c}}$ superconductivity. We therefore use the identified region as an empirical materials-space prior for candidate screening. Combining this criterion with the predicted $T_{\text{c}}$, we retrospectively recover the recently discovered nickelate superconductor La$_4$Ni$_3$O$_{10}$ and Nd${_{0.8}}$Sr$_{0.2}$NiO$_2$, and prioritize chemically distinct candidates, including Sr$_2$CoO$_4$ and Cs$_2$TaCl$_6$, for future investigation. These results establish a simple, data-driven materials-space criterion for navigating the search for high-$T_\mathrm{c}$ unconventional superconductors in the absence of a broadly applicable microscopic theory.

\section*{Results}

\subsection*{Feature engineering and dataset construction}
To construct a unified representation of superconducting materials, we encode each crystal structure into a fixed-length feature vector comprising three complementary components (Fig.~\ref{fig1}a). The first component captures compositional characteristics through elemental statistics. Specifically, we select 20 intrinsic elemental properties and calculate six statistical descriptors for each property, yielding a 120-dimensional composition-based vector~\cite{Ward2016}. The elemental properties and statistical descriptors are listed in the Supplementary Information~\cite{supp}. The second component consists of three structural descriptors directly obtained from the crystal structure: density, volume per atom, and packing fraction.

The third component is designed to capture material characteristics that are not explicitly encoded by simple compositional and structural descriptors. We therefore use latent representations from trained Atomistic Line Graph Neural Network (ALIGNN) models~\cite{Choudhary2021}. Rather than using the predicted physical properties themselves, we extract the latent representations from the final hidden layers of ALIGNN models as compact descriptors. Three regression models are trained using the Matbench dataset to predict the bulk modulus (KVRH), shear modulus (GVRH), and frequency of the last peak in the phonon density of states (Phono), respectively~\cite{Dunn2020}. Each model has a six-dimensional final hidden layer, providing a six-dimensional latent vector for the corresponding property. To incorporate information about the electronic states near the Fermi level, we additionally construct two electronic-structure datasets, DOS1 and DOS2, characterizing the presence of states near the Fermi level and their orbital character ($s$, $p$, $d$, and $f$), respectively~\cite{supp}. ALIGNN classification models are trained for these two tasks, yielding two- and four-dimensional output vectors corresponding to the respective classification schemes. The performance of all ALIGNN models is summarized in the Supplementary Information~\cite{supp}.

Because the original ALIGNN architecture is designed for stoichiometric structures, it cannot directly represent doped materials. We therefore modify the atomic embedding scheme for doped sites by replacing the embedding of the doped site with a composition-weighted average of the embeddings of its constituent atomic species. This modification enables a consistent representation of both stoichiometric and doped materials. Collectively, the three components provide complementary information on composition, crystal structure, and learned physical characteristics, resulting in a 147-dimensional feature vector for each material.

We next construct a curated superconducting dataset by combining experimentally reported superconducting properties with corresponding crystal structures (Fig.~\ref{fig1}b). Experimentally measured superconducting records from SuperCon~\cite{SuperCon2022} are mapped to crystal structures in the ICSD database~\cite{Zagorac2019} using the structure-matching procedure implemented in the 3DSC database~\cite{Sommer2023}. This procedure initially yields 86,445 structure--property entries. Because multiple structural entries can correspond to the same chemical composition, the resulting dataset contains substantial structural redundancy. After retaining only entries with $T_{\text{c}}\neq0$, 65,893 entries remain, corresponding to 6,711 unique chemical compositions. Treating all matched structures as independent samples could therefore disproportionately weight compositions represented by multiple structural variants and introduce bias into subsequent analyses.

To reduce this structural redundancy while retaining physically meaningful structural distinctions, we use the 147-dimensional feature vector to compare structures associated with the same chemical composition. Structures with highly similar feature vectors are consolidated into a single representative structure, whereas compositions whose matched structures exhibit substantially different representations are excluded (see Methods for details). This procedure yields a curated dataset of 4,253 superconducting materials with reduced structural redundancy, which we use for subsequent analyses.

Finally, because many of the 147 descriptors are strongly correlated, we group correlated features before performing the subsequent statistical analyses (Fig.~\ref{fig1}b). For composition- and structure-derived descriptors, hierarchical clustering based on pairwise feature correlations is used to merge highly correlated descriptors according to a predefined correlation threshold. For ALIGNN-derived representations, each learned property representation is retained as an individual feature group to preserve its associated physical context. For visualization and representative analyses, one descriptor is selected from each feature group to illustrate the characteristics of the corresponding feature category. The detailed clustering procedure and resulting feature groups are provided in the Methods and Supplementary Information, respectively~\cite{supp}.

\begin{figure*}[t]
\begin{center}
\includegraphics[width=5.5in, clip=true]{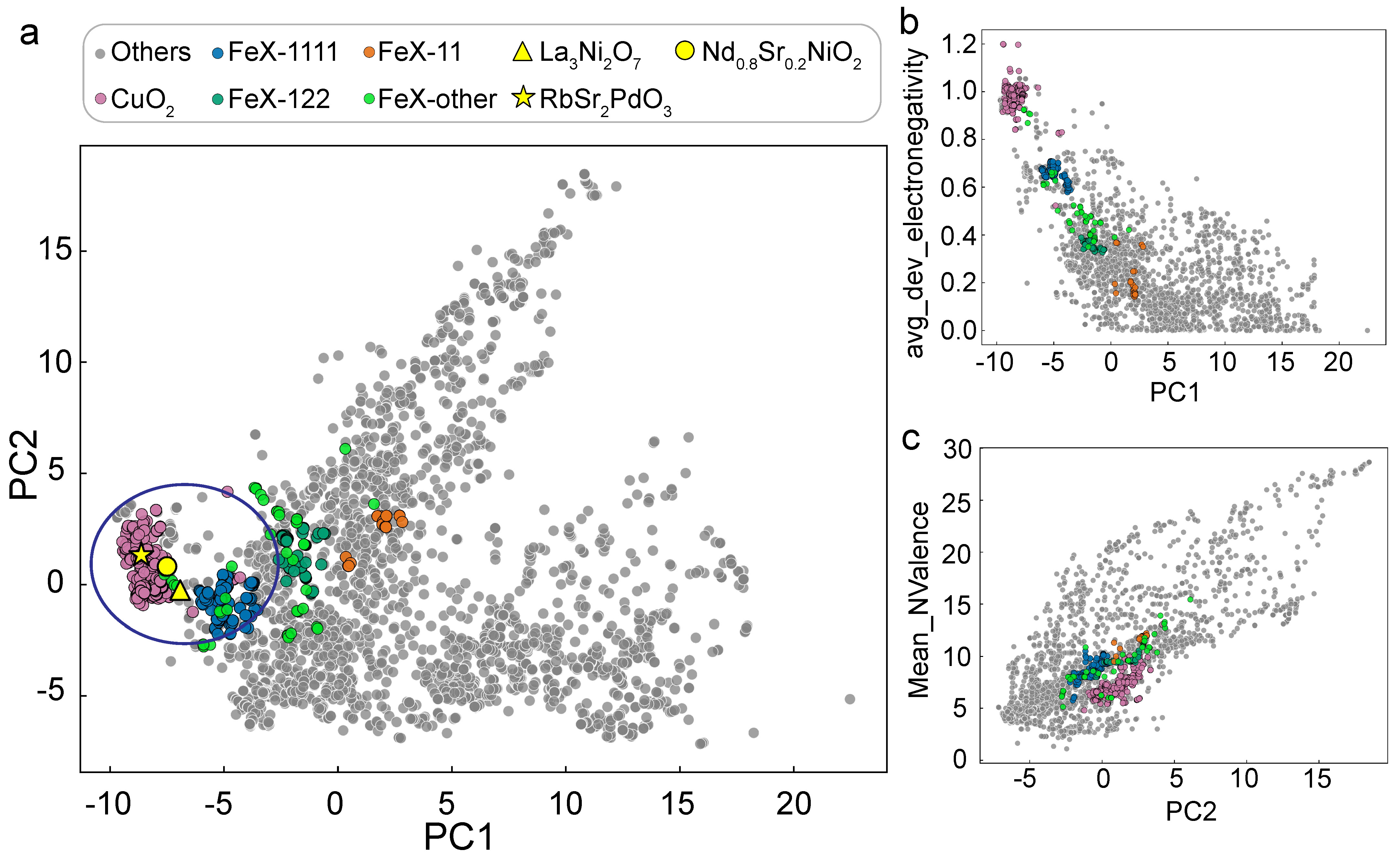}
\end{center}
\caption{\textbf{Unsupervised analysis of the superconducting materials space.} \textbf{a}, Projection of the curated superconducting dataset onto the PC1--PC2 plane. Cu-based and iron-based families concentrate in a common region (ellipse), into which La$_3$Ni$_2$O$_7$ (triangle), Nd$_{0.8}$Sr$_{0.2}$NiO$_2$ (circle) and RbSr$_2$PdO$_3$ (star), all of which are outside the training set, also fall. \textbf{b},\textbf{c}, Relationships between PC1 and \texttt{avg\_dev\_Electronegativity} (\textbf{b}) and between PC2 and \texttt{mean\_NValence} (\textbf{c}), with colors as in \textbf{a}.}
\label{fig2}
\end{figure*}

\subsection*{Unsupervised analysis}
To identify the dominant variations across the curated superconducting dataset, we perform principal component analysis (PCA) on the constructed feature representations. Although the correlated descriptors are grouped to facilitate feature organization and interpretation, PCA is performed on the complete 147-dimensional feature vectors rather than on the reduced set of representative descriptors. Prior to PCA, all feature vectors are standardized using statistics computed from a combined reference set comprising the superconducting materials and a large collection of structures from the Materials Project database~\cite{Jain2013}. This choice places superconducting materials on a common scale with the broader materials landscape and ensures that their positions in the resulting feature space can be directly compared with those of general materials. The resulting principal components therefore capture variation in the material representations independently of $T_{\text{c}}$.

As shown in Fig.~\ref{fig2}a, superconducting materials span a broad region of the PC1--PC2 plane. In contrast, the Cu- and Fe-based families concentrate in a common and relatively compact region, despite their distinct chemical compositions and crystal structures. This clustering suggests that these high-$T_{\mathrm{c}}$ unconventional superconductors share common characteristics in the constructed superconducting materials space. To examine whether this region extends beyond the materials used to construct the PCA representation, we further project three recently proposed superconducting materials that are not included in the training dataset onto the same PCA space. The experimentally confirmed Ni-based superconductors La$_3$Ni$_2$O$_7$~\cite{Sun2023} and Nd$_{0.8}$Sr$_{0.2}$NiO$_2$~\cite{LiDanfeng2019}, together with the theoretically predicted Pd-based superconductor RbSr$_2$PdO$_3$~\cite{Kitatani2023}, all fall within the region occupied by the Cu- and Fe-based families. Although these examples do not constitute a prospective validation, their independent projection suggests that the identified region captures characteristics shared across chemically distinct unconventional superconductor families.

To determine which physical descriptors are primarily responsible for this clustering, we calculate the correlations between the principal components and the clustered feature groups. PC1 shows its strongest correlation with the feature group represented by the average deviation of electronegativity (\texttt{avg\_dev\_Electronegativity}), whereas PC2 is primarily associated with the feature group represented by the mean number of valence electrons (\texttt{mean\_NValence}). The complete correlation analysis is provided in the Supplementary Information~\cite{supp}. As shown in Figs.~\ref{fig2}b and \ref{fig2}c, both descriptors exhibit clear correlations with their corresponding principal components, indicating that they capture the dominant variations within the constructed feature space. High-$T_{\mathrm{c}}$ unconventional superconductors preferentially occupy regions with relatively low PC1, corresponding to larger average electronegativity deviation, and intermediate PC2, corresponding to moderate mean valence-electron numbers.

These results identify a common materials-space signature of high-$T_{\mathrm{c}}$ unconventional superconductors,  characterized by a large average deviation of electronegativity together with an intermediate mean number of valence electrons. Other unconventional superconductors, such as heavy-fermion materials, occupy different regions of the feature space and generally have substantially lower transition temperatures~\cite{supp}. The characteristic region identified here therefore pertains specifically to the high-$T_{\mathrm{c}}$ unconventional superconducting families represented in our dataset. This signature is an empirical materials-space characteristic rather than a microscopic criterion for unconventional superconductivity. This constitutes the first main finding of our work.

\begin{figure*}[t]
\begin{center}
\includegraphics[width=6in, clip=true]{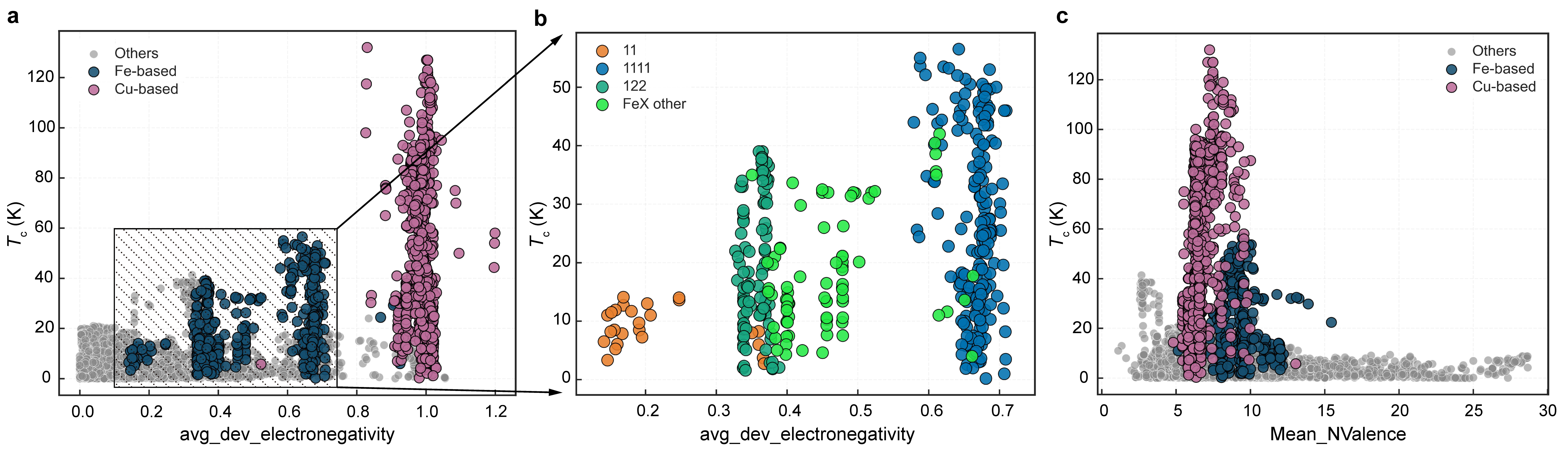}
\end{center}
\caption{\textbf{Dependence of $T_{\mathrm{c}}$ on the two leading descriptors.} \textbf{a}, $T_{\mathrm{c}}$ versus \texttt{avg\_dev\_Electronegativity}. The inset shows the iron-based subfamilies separately. \textbf{b}, $T_{\mathrm{c}}$ versus \texttt{mean\_NValence}.}
\label{fig3}
\end{figure*}

Having identified the characteristic materials-space region, we next ask how the observed $T_{\mathrm{c}}$ varies within this region. As shown in Fig.~\ref{fig3}a, the upper envelope of observed $T_{\mathrm{c}}$ increases with \texttt{avg\_dev\_Electronegativity}, indicating that larger electronegativity contrast is associated with higher observed $T_{\mathrm{c}}$ values. This trend persists within the iron-based family: the 11, 122, and 1111 subfamilies exhibit progressively larger \texttt{avg\_dev\_Electronegativity} together with progressively higher upper envelopes of observed $T_{\mathrm{c}}$ (inset of Fig.~\ref{fig3}a). This persistence within a single broad superconducting family suggests that the descriptor captures systematic variation beyond simply distinguishing different superconducting families. For the second descriptor, the upper envelope of observed $T_{\mathrm{c}}$ is highest when \texttt{mean\_NValence} lies in an intermediate range of approximately 7--8 (Fig.~\ref{fig3}b). Together, these results indicate that high $T_{\mathrm{c}}$ is favored within the identified materials space by increasing electronegativity deviation and an intermediate mean valence-electron number. This establishes the second main finding of our work: the two descriptors not only define a characteristic materials-space region for high-$T_{\mathrm{c}}$ unconventional superconductors, but also exhibit systematic correlations with the upper envelope of observed $T_{\mathrm{c}}$.

\begin{figure*}[t]
\begin{center}
\includegraphics[width=5.5in, clip=true]{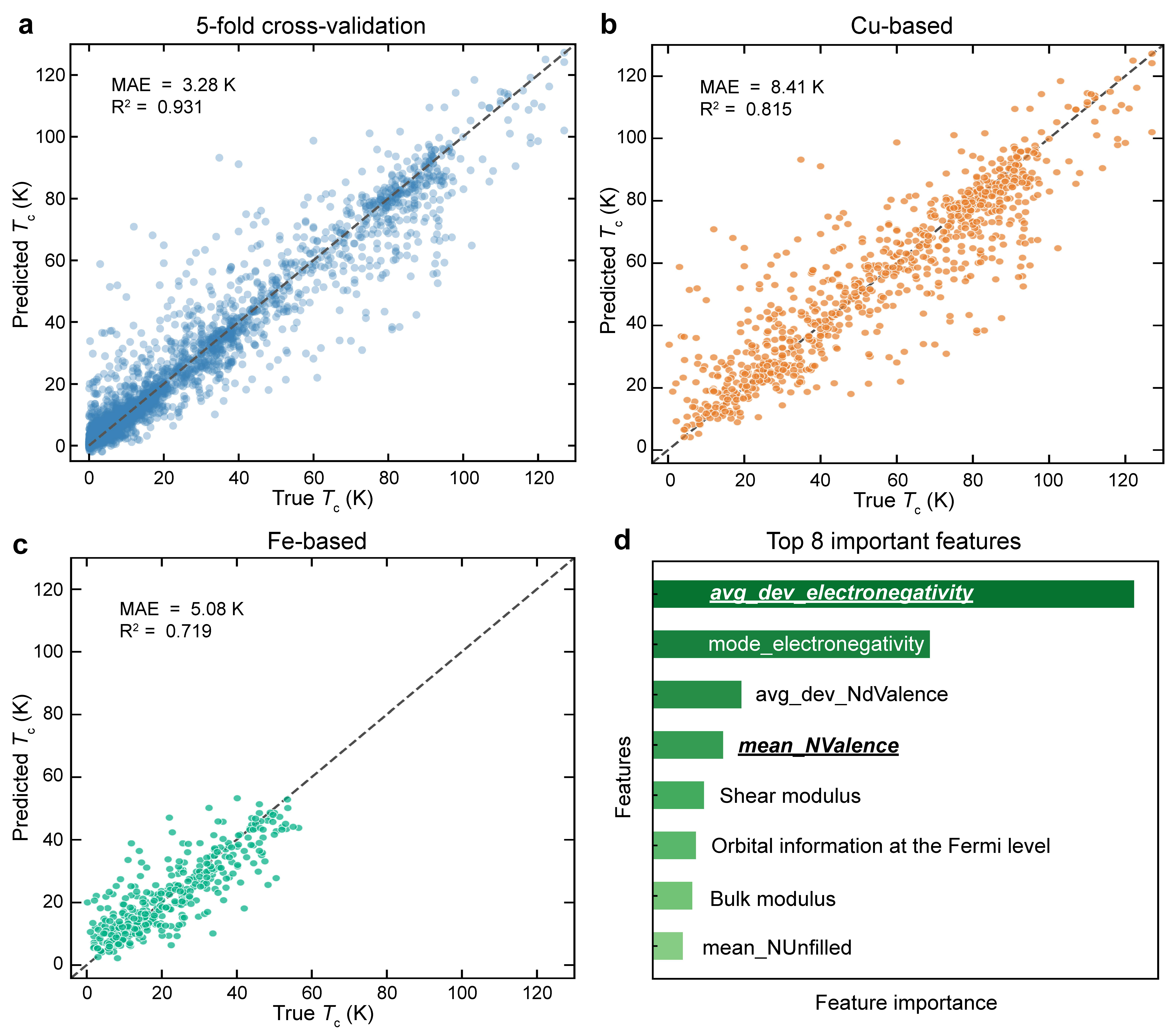}
\end{center}
\caption{\textbf{Supervised prediction of $T_{\mathrm{c}}$.} 
\textbf{a}, Predicted versus experimental $T_{\mathrm{c}}$ under five-fold cross-validation (MAE $=$ 3.28~K, $R^{2}=0.931$); the dashed line marks perfect prediction. 
\textbf{b,c}, Experimental and out-of-fold predicted $T_{\mathrm{c}}$ values for (b) cuprate and (c) iron-based superconductors obtained from five-fold cross-validation.  
\textbf{d}, The eight most important feature groups for $T_{\mathrm{c}}$ prediction.}
\label{fig4}
\end{figure*}

\subsection*{$T_{\mathrm{c}}$ prediction}

Although the unsupervised analysis identifies a characteristic materials-space region associated with high-\(T_{\mathrm{c}}\) superconductors, the two dominant descriptors revealed from this analysis capture the key trends defining this region but are insufficient by themselves for quantitative prediction of individual transition temperatures. We therefore perform supervised learning to predict \(T_{\mathrm{c}}\) based on the complete feature vector. We employ LightGBM~\cite{Ke2017}, a gradient-boosted decision-tree model, which is well suited to capturing nonlinear relationships among heterogeneous materials descriptors. The model is trained and evaluated using five-fold cross-validation on the curated superconducting dataset, with the hyperparameter selection and training procedure described in the Methods section. The resulting model achieves an average mean absolute error (MAE) of 3.284~K and an average coefficient of determination ($R^{2}$) of 0.931. As shown in Fig.~\ref{fig4}a, the predicted $T_{\mathrm{c}}$ values show good agreement with the experimentally measured values across the dataset. The model also maintains good predictive performance when evaluated separately on Cu-based and Fe-based superconductors in the held-out test sets (Figs.~\ref{fig4}b and \ref{fig4}c), demonstrating that its predictive power extends beyond distinguishing different superconducting families to resolving \(T_{\mathrm{c}}\) variations within each family. Ablation experiments using the same cross-validation scheme demonstrate that integrating structural descriptors and ALIGNN-based representations enhances the performance over composition-only features. Meanwhile, a modified ALIGNN model capable of processing doped structures is directly evaluated under the same protocol but remains inferior to the full-feature LightGBM model. Detailed model configurations and benchmark results are provided in the Supplementary Information~\cite{supp}.

Beyond predictive accuracy, the tree-based structure of LightGBM allows us to examine the relative importance of individual descriptors. As shown in Fig.~\ref{fig4}d, electronegativity-related descriptors dominate the feature importance ranking, with \texttt{avg\_dev\_Electronegativity} being the most important descriptor, followed by \texttt{mode\_Electronegativity}. Descriptors associated with valence electron number also contribute substantially, with \texttt{mean\_NValence} ranking fourth. Notably, these same descriptors were identified independently in the unsupervised analysis as the dominant factors associated with PC1 and PC2, respectively. The agreement between the unsupervised and supervised analyses therefore provides complementary evidence that average electronegativity deviation and mean valence-electron number capture robust information associated with high-$T_{\mathrm{c}}$ unconventional superconductors, rather than arising solely from the geometry of the PCA representation. A complementary SHapley Additive exPlanations (SHAP) analysis~\cite{lundberg2017unified,lundberg2020local} performed under the same cross-validation scheme yields a consistent ranking of the most important descriptor groups, further supporting the robustness of the feature-importance results. Larger electronegativity deviation also tends to contribute positively to the predicted $T_{\mathrm{c}}$, consistent with the increasing upper envelope of measured $T_{\mathrm{c}}$ (see Supplementary Materials for details~\cite{supp}).

The feature-importance analysis further shows that several ALIGNN-derived representations make substantial contributions to $T_{\mathrm{c}}$ prediction. In particular, the learned representations associated with shear modulus, electronic states near the Fermi level, and bulk modulus rank fifth, sixth, and seventh, respectively. Together with the ablation results, these findings show that the ALIGNN-derived structural and electronic representations provide information complementary to the simple elemental descriptors. Thus, whereas the two dominant descriptors capture the coarse organization of high-$T_\mathrm{c}$ unconventional superconductors in materials space, accurate $T_{\mathrm{c}}$ prediction at the level of individual compounds additionally depends on finer structural and electronic information.

\begin{figure*}[t]
\begin{center}
\includegraphics[width=5.5in, clip=true]{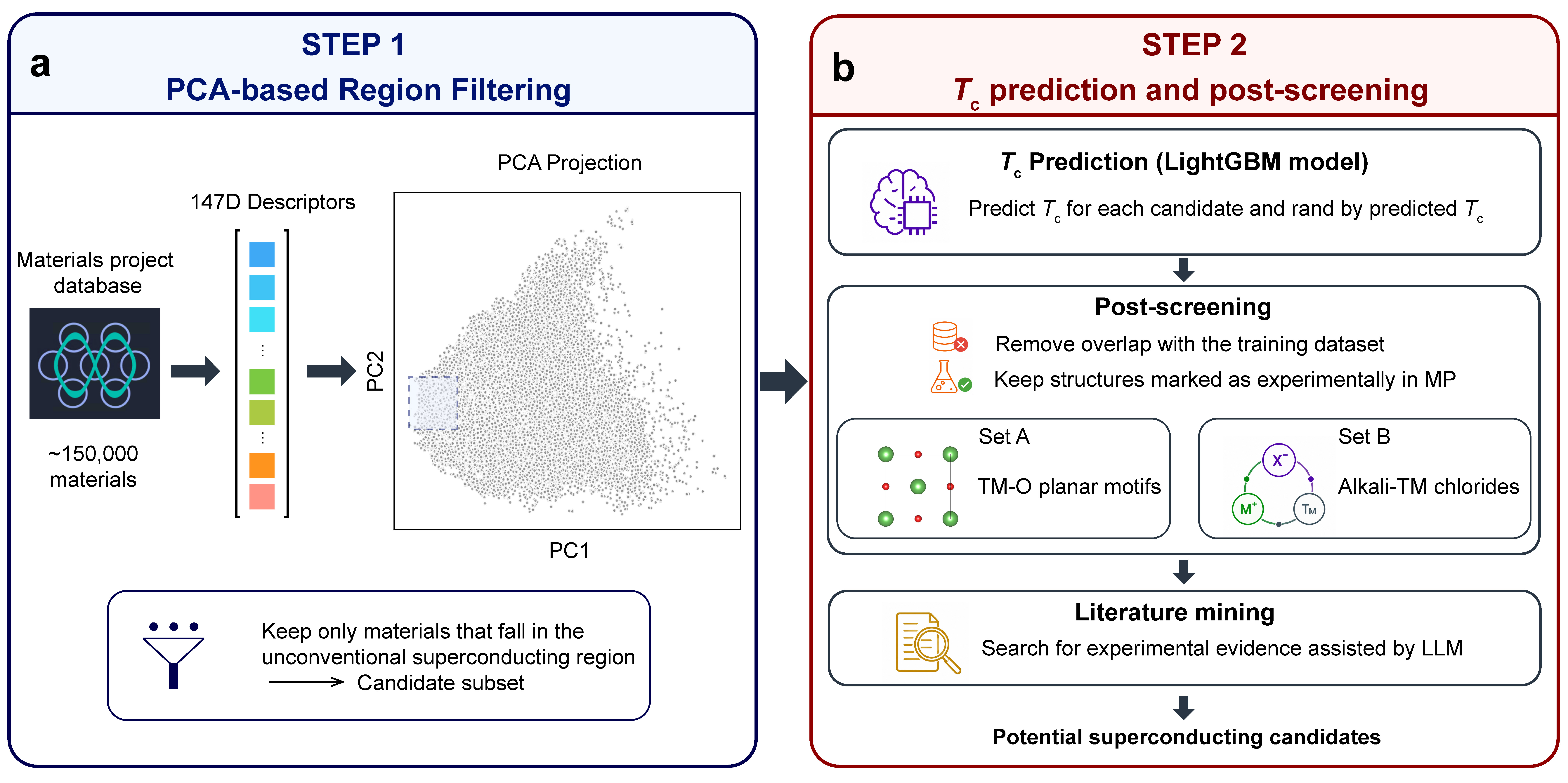}
\end{center}
\caption{\textbf{Two-step screening of potential superconducting candidates}. Materials from the Materials Project database are first filtered by their position in the PC1--PC2 plane, and the retained candidates are then ranked by predicted $T_{\mathrm{c}}$, post-screened into two targeted subsets (sets A and B), and examined through literature mining assisted by large language models.}
\label{fig5}
\end{figure*} 

\subsection*{Candidates screening}

Building on these results, we develop a two-step screening strategy that first restricts candidates to the characteristic region identified by PCA and then ranks them according to their predicted $T_{\mathrm{c}}$. A large collection of structures retrieved from the Materials Project database are first transformed into the constructed feature vectors and standardized using the same parameters derived from the combined Materials Project and superconducting datasets. These feature vectors are then projected onto the PC1–PC2 plane using the PCA transformation obtained from the superconducting dataset. As shown in Fig.~\ref{fig5}a, structures located within the characteristic region of high-$T_{\mathrm{c}}$ superconductors are selected as candidate materials.  For practical screening, we select materials from the Materials Project database satisfying $\mathrm{PC1}\leq -6.8$ and $-1\leq\mathrm{PC2}\leq3$. Using unconventional high-$T_{\mathrm{c}}$ superconductors ($T_{\mathrm{c}}>20$~K) in the superconducting dataset as the target class, 80.5\% of these materials fall within the selected PCA region, compared with 21.9\% of the full superconducting dataset, corresponding to an enrichment factor of approximately 3.7~\cite{supp}. This enrichment indicates that the selected region preferentially captures high-\(T_{\mathrm{c}}\) unconventional superconductors and therefore provides a useful criterion. These candidates are subsequently evaluated using the trained LightGBM model to obtain predicted $T_{\mathrm{c}}$ values and establish a ranking for further investigation.

\begin{table*}[htbp]
	\caption{\textbf{Representative materials prioritized by the ML-based screening workflow.} The table lists the Materials Project IDs, chemical formulas, and predicted superconducting transition temperatures ($T_{\mathrm{c}}$) for selected oxide candidates from Set A and chloride candidates from Set B.}
	\begin{center}
		\renewcommand{\arraystretch}{1.6}
		\begin{tabular*}{3.3in}
			{@{\extracolsep{\fill}}cccc}
			\hline
			\hline
			  & ID & Formula & Predicted $T_{\mathrm{c}}$ (K)\\
			\hline
			\multirow{5}{*}{set A}
			& mp-1077918 & Sr$_2$TcO$_4$ & 31.0 \\
			& mp-1205312 & Ca$_2$MnO$_4$ & 29.4 \\
			& mp-19102 & Sr$_2$FeO$_4$ & 24.3 \\
			& mp-18978 & Sr$_2$MnO$_4$ & 22.0 \\
			& mp-8672 & La$_2$PdO$_4$ & 17.1 \\
			\hline
			\multirow{5}{*}{set B}
			& mp-31363 & K$_2$TaCl$_6$ & 27.3 \\
			& mp-29400 & CsFeCl$_6$ & 24.6 \\
			& mp-568008 & Rb$_3$Mn$_2$Cl$_7$ & 24.0 \\
			& mp-28497 & Cs$_2$CrCl$_4$ & 23.5 \\
			& mp-28283 & CsTiCl$_3$ & 17.5 \\
			\hline
			\hline
		\end{tabular*}
	\end{center}
\end{table*}

After removing materials already included in the superconducting training dataset, we further restrict the candidates to structures marked as experimentally reported in the Materials Project database. The remaining candidates are ranked according to their predicted $T_{\mathrm{c}}$, and the top 20 materials are summarized in the Supplementary Information~\cite{supp}. These high-ranked candidates are mainly associated with either chemically complex compositions that remain largely unexplored for superconductivity or cuprate-related families absent from the training dataset. The latter trend reflects the dominance of Cu-based superconductors within the high-$T_{\mathrm{c}}$ region of the training data, causing the model to preferentially assign high predicted $T_{\mathrm{c}}$ values to chemically similar materials. This observation motivates additional targeted searches toward chemically simpler and more diverse candidate spaces.

To identify candidates with simpler compositions or potentially unexplored chemical characteristics, we perform two complementary searches. First, we restrict candidates to ternary compounds and search for materials containing transition-metal–oxygen (TM--O) planar motifs, motivated by the prevalence of CuO$_2$ planes in cuprate superconductors. We focus on cuprate-like motifs because the selected PCA region is more strongly associated with Cu-based superconductors. Using the geometric criteria described in the Methods section, we identify 43 candidate materials constituting set A. Second, to explore chemically distinct superconducting possibilities, we search for ternary alkali--transition-metal (A--TM) chlorides. This search is motivated by the strong association between larger average electronegativity deviation and higher attainable $T_{\mathrm{c}}$ revealed by our descriptor analysis. This procedure yields an additional set of 53 candidate materials constituting set B.

To further evaluate the superconducting relevance of these candidates, we performed systematic literature searches, assisted by a large language model for candidate identification, followed by manual verification against the original publications. For set A, the search identifies La$_4$Ni$_3$O$_{10}$ as a candidate with definitive experimental evidence of superconductivity under high pressure~\cite{Zhu2024}. Remarkably, the structure corresponds to the experimentally reported tetragonal $I4/mmm$ phase, and the predicted transition temperature of $6.8$~K falls within the experimentally observed $T_{\mathrm{c}}$ range~\cite{Zhu2024}. We further predict a $T_{\mathrm{c}}$ of $16.8$~K for Nd$_{0.8}$Sr$_{0.2}$NiO$_2$, which also lies within the selected region of the PCA plane and is in reasonable agreement with the experimentally reported transition temperature~\cite{LiDanfeng2019}. The successful retrospective identification of these recently established nickelate superconductors supports the effectiveness of the two-step screening strategy. Beyond these examples, several candidates show reported superconducting signatures under specific experimental conditions, while others remain unexplored. In contrast, no superconductivity-related reports were identified for the candidates in set B, suggesting that these chemically distinct compounds represent an unexplored materials space for future investigation. The complete literature-mining results are provided in the Supplementary Information~\cite{supp}, and representative candidates are highlighted in Fig.~\ref{fig1}c and Table~1.

\section*{Discussion}

In this work, we identify the average deviation of electronegativity and the mean number of valence electrons as two simple and interpretable descriptors that characterize the materials-space signature of high-$T_\mathrm{c}$ unconventional superconductors. Their importance is independently supported by both unsupervised analysis of the materials space and supervised prediction of transition temperatures. Guided by these descriptors, our two-step screening strategy successfully identifies the recently discovered nickelate superconductors La$_4$Ni$_3$O${_{10}}$ and Nd${_{0.8}}$Sr$_{0.2}$NiO$_2$, while also highlighting chemically distinct candidates for future investigation. These results demonstrate that interpretable machine learning can complement conventional materials discovery approaches by extracting useful design principles from experimentally established superconducting databases.

An important implication of our analysis concerns the choice of descriptors for quantitative prediction of unconventional superconductors. While the descriptors identified here effectively characterize the materials-space signature of high-$T_{\mathrm{c}}$ superconductors, they do not fully capture variations of $T_{\mathrm{c}}$ within individual superconducting families. In many unconventional superconducting systems, variations in $T_{\mathrm{c}}$ are strongly influenced by doping, whereas coarse composition-based descriptors remain nearly unchanged across a doping series. As shown in Fig.~\ref{fig3}a,b, doped compounds derived from a common parent phase can possess nearly identical descriptor values while exhibiting substantially different transition temperatures. Improving quantitative $T_{\mathrm{c}}$ prediction therefore requires descriptors that more directly encode doping-dependent electronic structures, carrier concentration, or related physical quantities. Furthermore, because pressure can strongly modify superconductivity in many materials, incorporating pressure information into superconductivity databases and using pressure as an explicit input feature may provide another route toward improving predictive performance.

Our present screening strategy is restricted to materials already catalogued in existing crystal-structure databases, limiting the accessible discovery space. A natural extension is therefore to combine the identified materials-space signature with generative models for crystal structures~\cite{Zeni2025, jiao2024space} to explore previously unknown compounds. Recent advances in reinforcement fine-tuning of pre-trained generative models enable optimization toward target properties~\cite{Park2026, cao2025, xu2026}, and the present results provide interpretable objectives for such optimization. Specifically, generated structures could be evaluated according to whether they occupy the characteristic materials-space region identified here and whether they exhibit high predicted $T_{\mathrm{c}}$ values from the supervised model.

Finally, the descriptors identified in this work should be viewed as empirical materials-space characteristics rather than microscopic criteria for unconventional superconductivity. A large average deviation of electronegativity suggests strong chemical contrast and charge transfer between constituent cation and anion sublattices, while an intermediate valence-electron number is associated with partially filled electronic states near the Fermi level, both of which are common features among several high-$T_{\mathrm{c}}$ unconventional superconducting families. However, establishing how these characteristics influence pairing interactions requires a microscopic theoretical framework beyond the present data-driven analysis. The descriptors identified here therefore provide experimentally grounded constraints and a materials-level perspective that may help bridge empirical discovery and future theories of unconventional superconductivity.

\section*{Methods}

\subsection*{Feature-based structural consistency filtering}

The initial superconducting dataset contains multiple candidate structures for identical chemical compositions due to the structure-matching procedure. To reduce structural redundancy and improve the reliability of the constructed superconducting materials dataset, we develop a feature-based structural consistency filtering strategy. For each chemical composition associated with multiple matched structures, the similarity among different structures is evaluated using both structural descriptors and ALIGNN-derived property representations.

First, a hard filtering step is applied based on the predicted electronic states near the Fermi level. Specifically, only structures with identical classification outputs from both DOS1 and DOS2 ALIGNN models are retained for subsequent analysis. 
DOS1 describes the presence or absence of electronic states near the Fermi level, whereas DOS2 characterizes their dominant orbital contributions ($s,p,d$ and $f$). This step removes matched structures with substantially different electronic characteristics before evaluating structural consistency.

For the remaining structures, a composite distance metric is constructed by integrating structural and property-based representations:
$D_{ij}=0.35||S_i-S_j||
+0.20d_{\mathrm{cos}}(G_i,G_j)
+0.20d_{\mathrm{cos}}(K_i,K_j)
+0.25d_{\mathrm{cos}}(P_i,P_j)$,
where $S$, $G$, $K$, and $P$ denote the structural representation and ALIGNN-derived representations associated with shear modulus, bulk modulus, and the frequency of the last peak in the phonon density of states, respectively. $||S_i-S_j||$ denotes the Euclidean distance between two structural representations, while $d_{\mathrm{cos}}$ denotes the cosine distance between two ALIGNN-derived property representations. The weighting coefficients are empirically chosen to integrate complementary structural and physical-property information.

For each chemical composition, the mean and maximum pairwise distances among all matched structures are calculated as $D_{\mathrm{mean}}$ and $D_{\mathrm{max}}$, respectively. Two threshold values, $t_{\mathrm{mean}}$ and $t_{\mathrm{max}}$, are introduced to control the structural consistency of the retained compositions. A composition is retained only when both criteria are satisfied:
$D_{\mathrm{mean}}<t_{\mathrm{mean}},
D_{\mathrm{max}}<t_{\mathrm{max}}$, where $t_{\mathrm{mean}}=0.1$ and $t_{\mathrm{max}}=0.2$ are used in this work. These criteria ensure that the retained chemical compositions exhibit consistent structural and property representations among different matched entries. For each retained composition, the structure with the smallest average distance to all other matched structures is selected as the representative structure. This procedure reduces the original dataset to 4,253 superconducting materials with reduced structural redundancy.

\subsection*{Feature correlation clustering}

To organize correlated descriptors and facilitate interpretation of the constructed feature space, hierarchical clustering is performed based on pairwise feature correlations. The Pearson correlation coefficient is calculated between every pair of descriptors, and the corresponding correlation distance is defined as:
$d_{ij}=1-|r_{ij}|$, where $r_{ij}$ denotes the Pearson correlation coefficient between descriptors $i$ and $j$.

Using the resulting distance matrix, hierarchical clustering is performed. Initially, each descriptor is treated as an individual cluster, and the distance between two clusters is calculated as the average pairwise distance between all descriptors belonging to the two clusters: $D(A,B)=\frac{1}{|A||B|}\sum_{i\in A}\sum_{j\in B}d_{ij}$, where $A$ and $B$ represent two feature clusters. The two closest clusters are iteratively merged until the predefined distance threshold is reached. In this work, a correlation distance threshold of 0.3 is adopted, corresponding to an absolute Pearson correlation coefficient of 0.7, to define groups of strongly correlated descriptors.

For the ALIGNN-derived representations, each learned property representation is treated as an individual feature group according to its corresponding physical meaning. The resulting feature groups are used only for organizing and interpreting correlated descriptors; all subsequent PCA and machine-learning analyses are performed using the complete 147-dimensional feature representation. The detailed clustering results are provided in the Supplementary Information~\cite{supp}.

\subsection*{Unsupervised and supervised learning}

To investigate the distribution of superconducting materials in the constructed feature space, PCA is performed on the standardized feature vectors. Before PCA projection, all descriptors are standardized using the mean and variance calculated 
from the combined dataset containing both superconducting materials and Materials Project structures. The principal components are obtained by eigenvalue decomposition of the covariance matrix and are used to visualize the materials-space distribution and identify dominant descriptor contributions.

For quantitative prediction of superconducting transition temperatures, a LightGBM regression model 
based on an ensemble of sequentially constructed decision trees is employed. The model is trained using the complete feature vectors and evaluated using five-fold cross-validation with shuffled data partitioning. For each fold, the model is trained with a maximum of 3000 boosting iterations and and an early-stopping criterion of 100 rounds based on the validation set. The learning rate is set to 0.06, with a maximum tree depth of 7 and 50 leaves per tree. To improve model generalization, both row and feature subsampling are applied, with a subsample ratio of 0.8 and a feature sampling ratio of 0.8. The model is optimized using MAE as the evaluation metric. Prediction performance is quantified using MAE and $R^2$. 

Feature importance is evaluated based on the gain contribution of each descriptor in the trained model. For each fold of the five-fold cross-validation, feature importance values are calculated independently from the corresponding trained model, and the final feature importance is obtained by averaging the values across all five folds.

\subsection*{Identification of TM--O planar environments}

To identify materials containing TM--O planar motifs analogous to CuO$_2$ planes, we develop a geometric structure analysis based on local coordination environments. For each TM site, neighboring oxygen atoms within a cutoff radius of 3.0$~\text{\AA}$ are collected. Candidate TM--O planes are identified by selecting oxygen quadruplets surrounding the TM center.

The local coordination environment is evaluated using three geometric criteria. First, the four oxygen atoms are required to exhibit similar TM--O bond lengths, quantified by either the maximum bond-length deviation or the relative standard deviation:
$\Delta d=d_{\max}-d_{\min}<0.25~\text{\AA}$,
or
$\sigma_d/\bar{d}<0.08$. Second, the TM atom and four surrounding oxygen atoms are required to be approximately coplanar. The best-fitting plane is obtained using singular value decomposition (SVD), and the maximum distance of atoms from this plane is required to be smaller than 0.20$~\text{\AA}$.
Finally, to identify planar oxygen atoms participating in extended TM--O networks, each oxygen atom is further examined for bridging configurations. An oxygen atom is considered a bridging oxygen when it is coordinated to at least two TM atoms within 2.4$~\text{\AA}$ and forms a nearly linear TM--O--TM bond angle larger than 150$^\circ$. A TM--O planar environment is identified when all four oxygen atoms in the candidate plane satisfy this bridging criterion.

Using these geometry-based screening, structures containing CuO$_2$-like TM--O planar motifs are identified for subsequent analysis. This approach serves as a geometry-based preliminary screening method and may also identify structures with similar local coordination environments but without genuine CuO$_2$-like planes. The screened candidates are further restricted to ternary compounds with elemental ratios corresponding to representative layered cuprate families, including 214, 327, 4310, and 112-type compositions. Candidate structures lacking genuine CuO$_2$-like planar motifs upon subsequent examination are excluded.

\section*{Acknowledgment}
J.W., H.X. and D.Q. disclose support for the research of this work from the Natural Science Foundation of China [grant No.~12350404], the National Key Research Program of China [grant No. 2025YFA1411400], the Quantum Science and Technology-National Science and Technology Major Project [grant No.~2021ZD0302600], the Science and Technology Commission of Shanghai Municipality [grant No.~23JC1400600 and No.~24LZ1400100], and the ``Shuguang Program'' supported by the Shanghai Education Development Foundation and Shanghai Municipal Education Commission.

J.W. conceived and supervised the project. D.Q. contributed to the development of the machine learning strategy by providing valuable insights into the supervised and unsupervised learning approaches. H.X. performed data curation, model training, data analysis, and materials screening. Y.Y. contributed to descriptor–$T_{\mathrm{c}}$ analysis and discussions on the materials-screening strategy. All authors contributed to the discussion of the results. H.X., D.Q. and J.W. wrote the manuscript with input from all authors. The authors declare no competing interests. Additional explanations and details regarding the model and generated materials can be found in the Supplementary Information.

\end{document}